\documentclass[12pt,a4paper]{article}
\pdfoutput=1
\usepackage{color}
\usepackage{amssymb,amsmath,bm,bbm}
\usepackage{epsf}
\usepackage{epsfig}
\usepackage{afterpage}
\usepackage{longtable}
\usepackage[dvipsnames]{xcolor}
\usepackage[linktoc=page,bookmarks=false,colorlinks=false,
linkbordercolor=RoyalBlue,citebordercolor=ForestGreen,urlbordercolor=CornflowerBlue]{hyperref}
\usepackage{latexsym,mathrsfs,dsfont}
\usepackage[normalem]{ulem} % for strikeout \sout
\usepackage[compress]{cite}
\usepackage{graphicx}
\usepackage{url}
\usepackage{paralist}
\usepackage{bbold}

\newcommand{\ord}{\mathcal{O}}

\newcommand{\RE}{{\rm Re}}

\newcommand{\tev}{\, {\rm TeV}}
\newcommand{\gev}{\, {\rm GeV}}

\newcommand{\vcb}{|V_{cb}|}

\newcommand{\vub}{|V_{ub}|}

\newcommand{\bsi}{B_6^{(1/2)}}
\newcommand{\bei}{B_8^{(3/2)}}

\def\epe{\varepsilon'/\varepsilon}
\newcommand{\beq}{\begin{equation}}
\newcommand{\eeq}{\end{equation}}
\newcommand{\be}{\begin{equation}}
\newcommand{\ee}{\end{equation}}
\newcommand{\bi}{\begin{itemize}}
\newcommand{\ei}{\end{itemize}}
\newcommand{\ba}{\begin{array}}
\newcommand{\ea}{\end{array}}
\newcommand{\beqa}{\begin{eqnarray}}
\newcommand{\eeqa}{\end{eqnarray}}
\newcommand{\bea}{\begin{eqnarray}}
\newcommand{\eea}{\end{eqnarray}}
\newcommand{\beqn}{\begin{eqnarray}}
\newcommand{\eeqn}{\end{eqnarray}}

\newcommand{\eps}{\epsilon}
\newcommand{\nn}{\nonumber}

\definecolor{red}{cmyk}{0,1,1,0.4}

\def\kpn{K^+\rightarrow\pi^+\nu\bar\nu}
\def\klpn{K_{L}\rightarrow\pi^0\nu\bar\nu}

\def\ksm{K_S\to\mu\bar\mu}

\def\klpll{K_L\to\pi^0\ell\bar\ell}

\newcommand{\kepe}{\kappa_{\varepsilon^\prime}}
\newcommand{\keps}{\kappa_{\varepsilon}}

\newcommand{\muNP}{{\mu_{\Lambda}}}
\newcommand{\muEW}{{\mu_{\rm ew}}}

\newcommand{\wc}[3][{}]{[{\cal C}_{#2}^{#1}]_{#3}}

\newcommand{\Wc}[2][{}]{{\cal C}_{#2}^{#1}}

\newcommand{\Op}[2][{}]{{\cal O}_{#2}^{#1}}

\usepackage{fancyhdr}
\advance \headheight by 3.0truept       % for 12pt mandatory...
\begin{document}

\begin{flushleft}
%{\em Version of \today}
%{\em Version of 14 September 2015}
\end{flushleft}

\vspace{-14mm}
\begin{flushright}
        {AJB-18-4}
\end{flushright}

\vspace{7mm}

\begin{center}
{\Large\bf
\boldmath{The Return of Kaon Flavour Physics}}
\\[8mm]
{\large\bf Andrzej~J.~Buras* \\[0.3cm]}
{\small 
      TUM Institute for Advanced Study, Lichtenbergstr.~2a, D-85748 Garching, Germany\\
Physik Department, TU M\"unchen, James-Franck-Stra{\ss}e, D-85748 Garching, Germany\\  E-mail: aburas@ph.tum.de }
\end{center}

\vspace{4mm}

\begin{abstract}
\noindent
Kaon flavour physics has played in the 1960s and 1970s a very important role in 
the construction of the Standard Model (SM) and in the 1980s and 1990s in SM tests  with the help of CP violation in  $K_L\to\pi\pi$ 
decays represented by $\varepsilon_K$ and the ratio $\epe$. In this millennium 
this role has been taken over by $B_{s,d}$ and $D$ mesons. However there is no doubt that in the coming years we will witness the return of kaon  flavour physics with the highlights being the measurements of the theoretically clean branching ratios for the rare decays $\kpn$ and $\klpn$ and the improved SM predictions for
the ratio $\epe$, for $\varepsilon_K$ and the $K^0-\bar K^0$ mixing mass difference $\Delta M_K$. Theoretical progress on 
 the decays  $K_{L,S}\to\mu^+\mu^-$ and $K_L\to\pi^0\ell^+\ell^-$ is also expected. They all are very sensitive to new physics (NP) contributions and  the correlations between them  should help us to identify new dynamics at very short distance scales. 
These studies  will be enriched when theory on the $K\to\pi\pi$ isospin amplitudes ${\rm Re} A_0$ and  ${\rm Re} A_2$ improves.
 This talk summarizes several aspects of this exciting field. In particular 
we emphasize the role of the Dual QCD approach in getting the insight into 
the numerical Lattice QCD results on $K^0-\bar K^0$ mixing and $K\to\pi\pi$ decays.
\end{abstract}

{\bf *} Talk given at the Epiphany Conference, Cracow, January 2018. To be published in the 
proceedings.

\setcounter{page}{0}
\thispagestyle{empty}
\newpage

\tableofcontents

\newpage

\section{Introduction}
A strategy for identifying new physics (NP) through flavour violating processes in twelve steps has been proposed in \cite{Buras:2013ooa}. Presently several of these steps cannot be realized, but this will certainly 
change in the 
coming years. In this talk I will concentrate 
on Kaon flavour physics and in particular on its main players:
\begin{itemize}
\item
$K^0-\bar K^0$ mixing with the parameter $\varepsilon_K$ representing mixing induced CP-violation in $K_L\to\pi\pi$ decays and  the 
$K_L-K_S$ mass difference $\Delta M_K$,
\item
The ratio $\epe$ representing the direct CP violation in $K_L\to\pi\pi$ decays
relative to the mixing induced one.
\item
The rare decays  $\kpn$ and $\klpn$, CP-conserving and CP-violating ones, respectively,
\item
Rare decays $K_{L,S}\to\mu^+\mu^-$ and $K_L\to\pi^0\ell^+\ell^-$, 
\item
The $K\to\pi\pi$ isospin amplitudes ${\rm Re} A_0$ and  ${\rm Re} A_2$ 
and in particular their ratio known under the name of the $\Delta I=1/2$ 
rule.  
\end{itemize}

During the last five years most of the flavour theorists concentrated their efforts on the explanation of the so-called $B$-physics anomalies in  $B\to K(K^*)\ell^+\ell^-$ 
 and $B\to D(D^*)\tau\nu_\tau$ decays. Several hundreds of papers were published
on them and numerous workshops have been organized to discuss possible NP 
behind them. Also this conference was dominated by these anomalies. While I took part in some of these discussions in the context 
of the so-called $P_5^\prime$ anomaly, I did not write a single paper on the
anomalies in the ratios $R_K$, $R_{K^*}$, $R_{D}$ and $R_{D^*}$. There 
were two reasons for it. First I am still not convinced that 
these four anomalies, related to the violation of lepton flavour universality, 
will survive the future more precise measurements, to be performed not only by the LHCb experiment but in particular by Belle-II. But I do hope very much 
that they will not disappear as they imply very interesting NP and moreover in 
view of the absence of 
direct signals for NP from ATLAS and CMS we need 
as many anomalies in flavour physics as possible.

The second  reason is
that having retired in 2012 I got slower and could simply not compete with 
much younger researchers in writing so many papers and definitely I did not
want to run behind a crowd which could be compared to the crowds 
 on the south-route to the summit of the Monte Everest. 
For emeriti  more pleasent is 
hiking in Norwegian mountaints, simply because one is basically almost alone 
meeting during the day only few tourists. If one looks at the number of 
papers written on Kaon flavour physics in the last five years, it is evident 
that working in this field is like hiking in  Norway. Therefore
 I changed my strategy and concentrated since 2014 \cite{Buras:2014maa}, with few exceptions, on Kaon flavour physics. A series of reviews on our work appeared 
in \cite{Buras:2014dxa,Buras:2015hna,Buras:2016qia,Buras:2016bch,Buras:2016egb}.

It is not the purpose of my talk to repeat all the material in these reviews
but rather concentrate on the 2017 news including also the very recent ones
which could not be presented at the Epiphany 2018 as the corresponding 
papers appeared just recently. However, my talk will in the first part concentrate on the main players, not necessarily in the order of their appearance  
in the list above. This will be the material of Section~\ref{sec:2}. 
In Section~\ref{sec:3} I will discuss some aspects of the work done in 
2017 and 2018. Finally, in Section~\ref{sec:4} I will present my shopping list
for the coming years.
\section{Main Players}\label{sec:2}
\boldmath
\subsection{The $\Delta I=1/2$ Rule}
\unboldmath
One of the puzzles of the 1950s was a large disparity between the measured 
values of the real parts of the isospin amplitudes $A_0$ and $A_2$ in $K\to\pi\pi$ decays,
which on the basis of usual isospin 
considerations were expected to be of the same order. In 2018 we know the 
experimental values of the real parts of these amplitudes very precisely 
\cite{Beringer:1900zz}
\be\label{N1}
{\rm Re}A_0= 27.04(1)\times 10^{-8}~\gev, 
\quad {\rm Re}A_2= 1.210(2)   \times 10^{-8}~\gev
\ee
and express
the so-called $\Delta I=1/2$ rule \cite{GellMann:1955jx,GellMann:1957wh}
\be\label{N1a}
{\rm R}=\frac{{\rm Re}A_0}{{\rm Re}A_2}=22.35.
\ee

In the 1950s QCD and  Operator Product Expansion did not exist and clearly 
one did not know that $W^\pm$ bosons existed in nature but using the ideas 
of Fermi \cite{Fermi:1934hr}, Feynman and Gell-Mann \cite{Feynman:1958ty} and Marshak and Sudarshan \cite{Sudarshan:1958vf} one could still 
 roughly estimate the amplitudes ${\rm Re}A_0$ and ${\rm Re}A_2$ to conclude 
that such a high value of ${\rm R}$ is a real puzzle. 

In modern times one can recover this puzzle by considering QCD in the large $N$ 
limit \cite{Buras:2014maa},  where $N$ is the number of colours. In this limit there are no QCD corrections to 
the Wilson coefficient of the current-current operator $Q_2 = (\bar su)_{V-A}\;(\bar ud)_{V-A}$ representing a simple tree-level $W^\pm$ exchange and the relevant hadronic matrix elements of this operator  can be calculated exactly in terms of pion 
decay constant $F_\pi$ and the masses $m_K$ and $m_\pi$ by just 
factorizing the operator matrix element into the product of matrix elements 
of quark currents. One finds then \cite{Buras:2014maa} 
\be\label{LO}
{\rm Re}A_0=3.59\times 10^{-8}\gev ,\qquad   {\rm Re}A_2= 2.54\times 10^{-8}\gev~, \qquad {\rm R}=\sqrt{2}
\ee
in plain disagreement with the data in (\ref{N1}) and (\ref{N1a}). 
It should be emphasized that the explanation of the  missing enhancement factor of $15.8$ in ${\rm R}$ through some dynamics must simultaneously give the correct values for ${\rm Re}A_0$ and  ${\rm Re}A_2$. 
This means that this dynamics should suppress  ${\rm Re}A_2$ by a factor of $2.1$, not more, and enhance ${\rm Re}A_0$ by a factor of $7.5$. This tells us 
that while the suppression of  ${\rm Re}A_2$  is an important ingredient in 
the $\Delta I=1/2$ rule, it is not the main origin of this rule.
 It is the enhancement of  ${\rm Re}A_0$. 

It should also be emphasized that the result in (\ref{LO}) has little to do with 
the so-called vacuum insertion approximation (VIA) but follows from 
the Dual QCD approach (DQCD) \cite{Bardeen:1986vp,Bardeen:1986uz,Bardeen:1986vz,Bardeen:1987vg} in which the factorization of matrix elements 
in question can be proven to be the property of QCD in the large $N$ limit
because in this limit  QCD at very low momenta becomes a free theory 
of mesons \cite{'tHooft:1973jz,'tHooft:1974hx,Witten:1979kh,Treiman:1986ep}.
 With  non-interacting mesons the factorization of matrix elements 
of four-quark operators into matrix elements of quark currents or quark 
densities is automatic.

The first step towards the explanation of the $\Delta I=1/2$ rule has 
been made through the pioneering 1974 calculations in \cite{Gaillard:1974nj,Altarelli:1974exa} where QCD renormalization group effects between $M_W$ and 
scales $\ord(1\gev)$,  to be termed {\it quark-gluon  evolution} in what follows, were done 
 at leading order in the renormalization group improved perturbation theory 
and now can be done at the NLO level. But if one continues to use 
hadronic matrix elements obtained by factorizing them the result is both 
scale and renormalization scheme dependent. Moreover as shown in \cite{Buras:2014maa}  the ratio $R$ is in the ballpark of $3-4$, certainly an improvement, 
but no explanation of its experimental value. In 1975 an attempt has 
been made to explain this rule by QCD penguins \cite{Shifman:1975tn} but 
in 1986 it was pointed out in the framework of DQCD that 
the current-current operators and not QCD penguins  are responsible dominantly for this rule \cite{Bardeen:1986vz}. This is obtained by performing 
a {\it meson evolution} from low scales at which factorization of matrix elements 
is valid in QCD to scales  $\ord(1\gev)$ at which the resulting matrix elements
are combined with their Wilson coefficients evaluated by the known renormalization group methods. As shown in \cite{Buras:2014maa} the pattern of meson evolution below $1\gev$ that includes also QCD penguins is similar to the one of quark-gluon evolution at short distance scales so that the matching between these two evolutions although not precise is acceptable. A good summary of the basic structure of  DQCD can be found in Sections 2 and 3 of  \cite{Buras:2014maa}.

The DQCD approach to weak decays developed in the 1980s has been improved 
in \cite{Buras:2014maa} through the inclusion of vector meson contributions in addition to pseudoscalars and improved through a better matching to short distance contributions.  Including 
QCD penguin contribution that at scales  $\ord(1\gev)$ amounts to a $10\%$ effect
in ${\rm Re}A_0$ one finds  \cite{Buras:2014maa}
\be\label{NLO+M+P}
{\rm Re}A_0\approx (17.0\pm 1.5)\times 10^{-8}\gev ,\qquad   {\rm Re}A_2\approx (1.1\pm 0.1)\times 10^{-8}\gev, \qquad {\rm R}\approx 16.0\pm 1.5~.
\ee
Even if the result for ${\rm Re}A_0$ is not satisfactory, it should 
be noted that the QCD dynamics identified by us was able to enhance 
the ratio ${\rm R}$ by an order of magnitude. We therefore conclude that 
QCD dynamics is dominatly responsible for the $\Delta I=1/2$ rule. The 
remaining piece in ${\rm Re}A_0$ could come from final state interactions (FSI) 
between pions as advocated in \cite{Antonelli:1995gw,Bertolini:1995tp,Pallante:1999qf,Pallante:2000hk,Buchler:2001np,Buchler:2001nm,Pallante:2001he} bringing the values of R in (\ref{NLO+M+P}) closer to its experimental value. 
Some support for this claim  comes from the recent reconsideration of the role of FSI in the $\Delta I=1/2$ rule in
\cite{Buras:2016fys}. As investigated in \cite{Buras:2014sba} also NP could enter at some level. 

After heroic efforts over many years also lattice QCD by means of 
 very sophisticated and tedious numerical calculations
 made impressive progress towards the 
explanation of the $\Delta I=1/2$ rule within the SM.
The most recent result from the RBC-UKQCD collaboration reads \cite{Bai:2015nea}
\be\label{DRULEL}
\left(\frac{{\rm Re}A_0}{{\rm Re}A_2}\right)_{{\rm lattice~QCD}}=31.0\pm11.1 \,
\ee
and in agreement with the 1986 result from DQCD \cite{Bardeen:1986vz} also this result is governed by current-current operators. But the uncertainty is still very large and it 
will be interesting to see whether lattice will be able to come closer to the
data than it is possible using DQCD. One should also stress that the 
lattice value for ${\rm Re}A_2$ has a much smaller error than ${\rm Re}A_0$
 and agrees well with  the data.

To summarize, from my point of view the dominant dynamics behind the 
$\Delta I=1/2$ rule has been identified within the DQCD approach already 
in 1986 \cite{Bardeen:1986vz} and has been confirmed through improved calculations in 2014
\cite{Buras:2014maa}. This dynamics is very simple. It is just short distance 
(quark-gluon) evolution of current-current operators down to scales  $\ord(1\gev)$ 
followed by meson evolution down to scales $\ord(m_\pi)$ at which the hadronic 
matrix elements factorize and can easily be calculated. I doubt that 
the remaining piece can be fully explained by NP as this would lead to
a very large fine-tuning in $\Delta M_K$ as demonstrated in \cite{Buras:2014sba}. It is likely that FSI and additional subleading corrections not included 
in the result in (\ref{NLO+M+P}) could be responsible for
 the missing piece. However I do not think that the present analytic 
methods like DQCD or the methods advocated by Pich and collaborators, as reviewed recently in \cite{Gisbert:2017vvj}, are sufficiently powerful 
to answer the question at which level NP enters the amplitudes 
${\rm Re}A_0$ and ${\rm Re}A_2$.  Here lattice QCD should provide valuable
answers and I am looking forward to improved results on these two amplitudes 
from RBC-UKQCD collaboration and other lattice groups. This would 
provide two additional important constraints on NP models.
\boldmath
\subsection{$\varepsilon_K$ and $\Delta M_K$}
\unboldmath

 The parameter $\varepsilon_K$ and the $K_L-K_S$ mass difference have played 
already for decades an important role in the constraints on NP. There is some tendency that $\varepsilon_K$ in the SM is below the data 
\cite{Lunghi:2008aa,Buras:2008nn,Bona:2006ah,Charles:2015gya},
but certainly one cannot talk presently about
an anomaly in $\varepsilon_K$. Indeed this  depends on whether inclusive or exclusive determinations of $\vub$ and $\vcb$ are used and with the inclusive ones SM value of $\varepsilon_K$ agrees well with the data. But then as emphasized in \cite{Blanke:2016bhf} $\Delta M_s$ and $\Delta M_d$ are significantly above the data. Moreover, this is true  for the whole class of  CMFV models. 
Related discussions can be found  in \cite{Buras:2013raa,Buras:2014sba,Bailey:2015frw,Bazavov:2016nty,DiLuzio:2017fdq}. This tension increased recently due to the improved 
lattice calculations \cite{Bazavov:2017lyh} and could signal new complex phases beyond 
the CKM 
phase as only such phases could decrease $\Delta M_s$ and $\Delta M_d$ through 
destructive intereference between SM and NP contributions. 

Such new phases could have an impact not only on $\varepsilon_K$ but as 
emphasized in  \cite{Buras:2015jaq} also on $\Delta M_K$. 
The point is that $\Delta M_K$ is proportional to the real part of a square
of a complex coefficient $C_K$ and a new phase modyfing its imaginary part will
quite generally decrease the value of $\Delta M_K$ relative to the SM 
estimate simply because
\be
 (\Delta M_K)^{\text{NP}} =  c \left[({\rm Re}\,C_K)^2-({\rm Im}\,C_K)^2\right]
\ee
with $c$ being positive. The uncertainty in the SM estimate of $\Delta M_K$ is unfortunately still very large  \cite{Brod:2011ty} so that we cannot presently decide whether a positive or negative NP contribution to $\Delta M_K$ if any is required. Future lattice QCD calculations of long distance contributions to $\Delta M_K$ could help in this respect  \cite{Bai:2014cva,Christ:2015pwa}. In 
DQCD they are found to amount to $20\pm 10\%$  of the measured $\Delta M_K$  \cite{Bijnens:1990mz,Buras:2014maa}. In the case of  $\varepsilon_K$ 
such long distance contributions to $\varepsilon_K$ are
 below $10\%$ and have been reliably calculated in \cite{Buras:2008nn,Buras:2010pza}.

Now if NP contributes significantly to $\varepsilon_K$ and $\Delta M_K$, one 
has to consider new local operators in addition to the SM operator so 
that the full operator basis is given  as follows
\cite{Gerard:1984bg,Gabbiani:1996hi}
\bea \label{eq:susy}
\mathcal{O}_1 &=& \bar s^\alpha \gamma_\mu P_L d^\alpha \ \bar s^\beta 
\gamma_\mu P_L 
d^\beta\, , \nn \\ 
\mathcal{O}_2 &=& \bar s^\alpha P_L d^\alpha \ \bar s^\beta P_L d^\beta 
\, , \nn \\ 
\mathcal{O}_3&=& \bar s^\alpha P_L d^\beta \ \bar s^\beta P_L  d^\alpha 
\, , \\ 
\mathcal{O}_4 &=& \bar s^\alpha P_L d^\alpha \ \bar s^\beta P_R d^\beta 
\, , \nn \\ 
\mathcal{O}_5&=& \bar s^\alpha P_L d^\beta \ \bar s^\beta P_R d^\alpha
\, , \nn 
\eea 
with $\alpha,\beta$ being colour indices and $P_{R,L}=(1\pm\gamma_5)/2$. Only 
$\mathcal{O}_1$ is present in the SM. Moreover also operators with 
$P_L$ and $P_R$ interchanged contribute.

The 
Wilson coefficients of these operators have been known at the NLO level 
 \cite{Ciuchini:1997bw,Buras:2000if} already for almost two decades. Recently 
also significant progress in the evaluation of $K^0-\bar K^0$ matrix elements 
by  ETM, SWME and RBC-UKQCD lattice collaborations \cite{Carrasco:2015pra,Jang:2015sla,Garron:2016mva,Boyle:2017skn,Boyle:2017ssm} has been made.

It is customary to represent the results for the $K^0-\bar K^0$ matrix 
elements of the operators in question in terms of $B_i$ parameters. 
In the vacuum insertion approximation (VIA) they are simply given by
\be\label{VIA}
B_1=B_2=B_3=B_4=B_5=1 \,\qquad {(\rm VIA)}
\ee
and moreover do not depend on the renormalization scale $\mu$ as predicted by 
QCD. 
Already this property of VIA, which is based on the factorization of matrix elements
of four-quark operators into products of quark currents or quark densities, 
is problematic as generally these parameters depend on 
$\mu$. 

Now RBC-UKQCD collaboration working at $\mu=3\gev$ finds
\cite{Garron:2016mva,Boyle:2017skn,Boyle:2017ssm}
\be\label{L23}
B_1=0.523 (9)(7), \qquad B_2=0.488(7)(17), \qquad  B_3=0.743(14)(65)
\ee
and
\be\label{L45}
B_4=0.920(12)(16),\qquad B_5=0.707(8)(44),
\ee
with the first error being statistical and the second systematic. Similar 
results are obtained by EMT and SWME collaborations although the values for 
$B_4$ and $B_5$ from the ETM collaboration are visibly below the ones from 
given above: $B_4=0.78(4)(3)$ and $B_5=0.49(4)(1)$.
Except 
for $B_4$ all values differ significantly from unity prohibiting the use 
of VIA. 

To our knowledge no lattice group made an attempt to understand this 
pattern of values, probably because  within lattice QCD which works 
at scales $\ord(2-3)\gev$ this pattern cannot
be understood. On the other hand it has been recently demonstrated in {\cite{Buras:2018lgu}} that this pattern can be
understood within DQCD approach because in this approach an insight in 
the QCD dynamics at very low scales up to $ 1\gev$ can be obtained through 
meson evolution followed by the usual RG QCD evolution as already discussed 
above in the context of the $\Delta I=1/2$ rule.

The case of $B_1$ is well known. In the large $N$ limit one finds $B_1=3/4$ 
\cite{Buras:1985yx}.
The meson evolution followed by quark-gluon evolution brings it in the ballpark of the lattice result in (\ref{L23}). In this particular case
one usually multiplies the result
 by the corresponding SD  renormalization group factor to find the scale and renormalization scheme independent  $\hat B_K=0.73\pm0.02$  \cite{Buras:2014maa} in a very good 
agreement with the  world average of lattice QCD calculations  
$\hat B_K=0.766\pm 0.010$ \cite{Aoki:2016frl}. 

In the case of the BSM operators $\mathcal{O}_i$ with $i=2-5$ the construction 
of scale independent $\hat B_i$ parameters, although possible, is not particular
useful because  $\mathcal{O}_2$ mixes under renormalization with $\mathcal{O}_3$
and $\mathcal{O}_4$ with  $\mathcal{O}_5$. This mixing is known at the NLO level
 \cite{Ciuchini:1997bw,Buras:2000if} and useful NLO expressions for $\mu$ dependence of hadronic matrix elements and their Wilson coefficients can be found in 
\cite{Buras:2001ra}. 

In the large $N$ limit one finds {\cite{Buras:2018lgu}}
\be\label{B25}
B_2=1.20, \qquad  B_3=3.0\,, \qquad 
B_4=1.0, \qquad  B_5=0.2 \qquad ({\rm large~N~limit})\,.
\ee

These results differ significantly from lattice results but apply 
to $\mu=\ord(m_\pi)$ while the lattice results where obtained at $\mu=3\gev$.
It is therefore remarkable that the pattern  
\be\label{pattern1}
B_2 < B_5\le B_3 < B_4 \, \qquad (\mu= 3\gev)~~~({\rm Lattice~QCD})
\ee
can indeed be understood within DQCD although there one
finds first
\be\label{pattern2}
B_5 < B_4 < B_2 < B_3\, \qquad (\mu= \ord(m_\pi))~~({\rm DQCD})\,.
\ee

As meson evolution with the inclusion of pseudoscalar mesons can be done only up to $\mu=0.65\pm 0.05\gev$ let us use the standard 
RG equations to find first lattice values for $B_i$ at $\mu=1\gev$, where 
perturbation theory is still reliable. From central values in 
(\ref{L23}) and (\ref{L45}) one 
finds at $\mu=1\gev$ {\cite{Buras:2018lgu}.}
\be
B_2=0.608,\quad B_3=1.06, \quad B_4=0.920, \quad B_5=0.519 ~~~({\rm Lattice~QCD}).
\ee
Using ETM values for $B_4$ and $B_5$ one would find $B_4=0.78$ and $B_5=0.24$.

We observe that $B_2$, $B_3$ and $B_5$, all moved towards their large $N$ 
values in (\ref{B25})  while $B_4$  did not change in LO approximation. These results are already very encouraging.
The rest of the job is done by meson evolution. Starting with the values in 
 (\ref{B25})  and performing meson evolution in the chiral limit   one finds 
at order $1/N$ {\cite{Buras:2018lgu}}
\be\label{R1}
B_2(\Lambda)=1.2\, \left[1 - \frac{8}{3}\, \frac{\Lambda^2}{(4\pi F_K)^2}\right],
\qquad
B_3(\Lambda)=3.0\, \left[1 - \frac{16}{3}\,  \frac{\Lambda^2}{(4\pi F_K)^2} \right],
\ee
\be\label{R3}
B_4(\Lambda)=1.0\, \left[1 -  \frac{4}{3}\,  \frac{\Lambda^2}{(4\pi F_K)^2}\right],
\qquad
B_5(\Lambda) = 0.23\,
\left[1 + 4\,  \frac{\Lambda^2}{(4\pi F_K)^2}\right]\,\,,
\ee
where $\Lambda$ is the cut-off of DQCD which allows us to separate the non-factorizable meson evolution from the quark-gluon one. 
The general trend already observed in the quark-gluon evolution is nicely outlined in the meson evolution with a strong suppression of $B_2$, 
an even stronger suppression of $B_3$, a smooth evolution of $B_4$ and 
a strong enhancement of $B_5$.

Consequently for 
 $\Lambda=0.7\gev$ one finds
\be
B_2=0.79,\quad B_3=0.96, \quad B_4=0.83, \quad B_5=0.30 ~~~({\rm DQCD}).
\ee
We note also that the values for $B_4$ and $B_5$ are in between those from 
RBC-UKQCD and ETM collaborations and we are looking forward to new improved lattice 
results for all four parameters in order to see how well DQCD reproduces 
LQCD numbers in question.

In any case as the meson evolution has been performed in the chiral limit without
the inclusion of vector meson contributions this result should be considered as
not only satisfactory but remarkable as  our calculations involved only one parameter, the 
cut-off scale $\Lambda$ which in any case should be around $0.7\gev$ if only 
pseudoscalar meson contributions are taken into account. It demonstrates the importance of the QCD dynamics at scales 
below $1\gev$ and gives additional support to our claim that meson evolution 
 is the dominant QCD dynamics responsible for the $\Delta I=1/2$ rule.

We are not aware of any analytical approach that could provide such 
insight in lattice QCD results in question. We challange the chiral perturbation theory experts to provide an insight into the values of $B_i$ from LQCD in their framework, in particular without using lower energy constants obtained from 
LQCD.

\boldmath
\subsection{$\kpn$ and $\klpn$}
\unboldmath
These two very rare decays are exceptional in the flavour physics as 
their branching ratios are know for fixed CKM parameters within an 
uncertainty of $2\%$ which to my 
knowledge cannot be matched by any other meson decay.  Indeed, they are theoretically 
very clean and their branching ratios have been calculated within  the SM including NLO QCD corrections to the top quark contributions 
\cite{Buchalla:1993bv,Misiak:1999yg,Buchalla:1998ba}, 
 NNLO QCD corrections  to the charm contribution in $\kpn$ \cite{Gorbahn:2004my,Buras:2005gr,Buras:2006gb} and  NLO electroweak corrections \cite{Brod:2008ss,Brod:2010hi,Buchalla:1997kz}.
Moreover, extensive calculations of isospin breaking effects and 
non-perturbative effects have been done \cite{Isidori:2005xm,Mescia:2007kn}. 
Therefore, once the CKM parameters $\vcb$, $\vub$ and $\gamma$  will be 
precisely determined in tree-level decays, these two decays will 
offer  excellent tests of the SM and constitute  very powerful probes of NP. Reviews of these two decays can be found in 
\cite{Buras:2004uu,Komatsubara:2012pn,Buras:2013ooa,Blanke:2013goa,Smith:2014mla}. In particular in \cite{Buras:2001af} bounds on $K\to\pi\nu\bar\nu$ decays 
in correlation with the unitarity triangle and $\sin 2\beta$ within models with
minimal flavour violation have been derived. See also interesting recent papers 
of the impact of lepton flavour non-universality on these decays 
\cite{Crivellin:2016vjc,Bordone:2017lsy,Fajfer:2018bfj} and 
right-handed neutrinos \cite{He:2018uey}.

It is really exciting that after twenty five years of waiting \cite{Buchalla:1993bv,Buras:1994ec}, the prospects of measuring the 
branching ratios for these two {\it golden} modes with good precision within the next five years are very good. Indeed,
the NA62 experiment at CERN  has recently found one event of $\kpn$ decay and twenty SM-like events are expected until the end of 2019. Eventually NA62 expects to measure the $\kpn$ branching ratio with the precision of $\pm10\%$ \cite{Rinella:2014wfa,Romano:2014xda}. Also the KOTO experiment at J-PARC should make a significant progress in measuring the branching ratio for $\klpn$ \cite{Komatsubara:2012pn,Shiomi:2014sfa}. 

Here it will suffice to quote parametric expressions 
for  branching ratios  $\mathcal{B}(\kpn)$ and $\mathcal{B}(\klpn)$ in the SM
in terms of the CKM inputs \cite{Buras:2015qea}  
\begin{align}
    \mathcal{B}(\kpn) = (8.39 \pm 0.30) \times 10^{-11} \cdot
    \bigg[\frac{\left|V_{cb}\right|}{40.7\times 10^{-3}}\bigg]^{2.8}
    \bigg[\frac{\gamma}{73.2^\circ}\bigg]^{0.74},\label{kplusApprox}
\end{align}
\begin{align}
    \mathcal{B}(\klpn) = (3.36 \pm 0.05) \times 10^{-11} \cdot
    &\bigg[\frac{\left|V_{ub}\right|}{3.88\times 10^{-3}}\bigg]^2
    \bigg[\frac{\left|V_{cb}\right|}{40.7\times 10^{-3}}\bigg]^2
    \bigg[\frac{\sin(\gamma)}{\sin(73.2^\circ)}\bigg]^{2}. \label{k0Approx}
    %&\times\bigg[\frac{\left|V_{us}\right|}{0.2252}\bigg]^{-2}
    %\bigg[\frac{\sin(\gamma)}{\sin(70^\circ)}\bigg]^{2}.
\end{align}
The parametric relation for $\mathcal{B}(\klpn)$ is exact, while for $\mathcal{B}(\kpn)$ it gives an excellent approximation:
for the large ranges $37 \leq |V_{cb}|\times 10^{3} \leq 45$ and $60^\circ \leq \gamma \leq 80^\circ$ it is accurate to 1\% and 0.5\%, respectively. The exposed  errors are non-parametric ones. They originate in the left-over uncertainties
in QCD and electroweak corrections and other small uncertainties. For $\kpn$ the error is larger due to the relevant charm contribution that can be neglected 
for $\klpn$. In the case of $\mathcal{B}(\kpn)$ we have absorbed $|V_{ub}|$ into the non-parametric error due to the weak dependence on it. 

The virtue of these formulae is that they allow easily to monitor the changes in the values of branching ratios in question, which clearly will still take place before the values on $\vcb$, $\vub$ and $\gamma$ from tree-level decays will be precisely known. The error budgets can be found in Fig.~1 of \cite{Buras:2015qea}. They tell us, as already inferred from (\ref{kplusApprox}) 
and (\ref{k0Approx})  that for $\kpn$ the crucial CKM element is
$\vcb$ and for $\klpn$ all three: $\vcb$, $\vub$ and $\gamma$.

Using (\ref{kplusApprox}) and  (\ref{k0Approx}) together with an average 
provided in  \cite{Buras:2015qea}
\be\label{avg}
\vcb_\text{avg}=(40.7\pm1.4)\cdot 10^{-3},\qquad \vub_\text{avg}=(3.88\pm0.29)\cdot 10^{-3}.
\ee
one finds with $\gamma = (73.2^{+6.3}_{-7.0})^\circ$ 
\begin{align}\label{PREDA}
    \mathcal{B}(\kpn) &= \left(8.4 \pm 1.0\right) \times 10^{-11}, \\
    \mathcal{B}(\klpn) &= \left(3.4\pm 0.6\right) \times 10^{-11}.    
\end{align}
While the values in (\ref{avg}) will change in time, we expect that both 
branching ratios will not be modified by more than $15\%$ and the errors will 
be reduced significantly due to better determination of  $\vcb$, $\vub$ and $\gamma$.

 Experimentally we have \cite{Artamonov:2008qb}
\be\label{EXP1}
\mathcal{B}(\kpn)_\text{exp}=(17.3^{+11.5}_{-10.5})\cdot 10^{-11}\,,
\ee
and very recently NA62 collaboration observing one event quotes 
\be\label{EXP1a}
\mathcal{B}(\kpn)_\text{exp}=(28^{+44}_{-23})\cdot 10^{-11}\,,\qquad ({\rm NA62}).
\ee
This result should be improved in 2019.
The $90\%$ C.L. upper bound on $\klpn$  reads \cite{Ahn:2009gb} 
\be\label{EXP2}
\mathcal{B}(\klpn)_\text{exp}\le 2.6\cdot 10^{-8}\,.
\ee
It should also be improved by KOTO in the coming years.

\boldmath
\subsection{$\epe$ Striking Back}
\unboldmath
One of the stars of flavour physics in the 1990s  was the ratio
$\epe$  that measures the size of the direct CP violation in $K_L\to\pi\pi$ 
relative to the indirect CP violation described by $\varepsilon_K$. On the 
experimental side the world average from NA48 \cite{Batley:2002gn} and KTeV
\cite{AlaviHarati:2002ye,Abouzaid:2010ny} collaborations reads
\be\label{EXP}
(\epe)_\text{exp}=(16.6\pm 2.3)\times 10^{-4} \,.
\ee

On the theory side a long-standing challenge in  making predictions for $\epe$ within the SM  has been the significant cancellation  of QCD penguin contributions by electroweak penguin contributions to this ratio. In the SM, QCD penguins
give a positive contribution and electroweak penguins a negative one. In the 
1980s, when the mass of the top quark was not known and $m_t$ in the ballpark
of $50-100\gev$ has been used in the analyses of $\epe$, electroweak penguin 
contributions governed by $Z^0$-penguins could be neglected and only 
QCD penguins and isospin breaking corrections were taken into account. The 
SM prediction was then close to the one in (\ref{EXP}) \cite{Buras:1987qa}. 
The situation changed in 1989 when it was demonstrated in \cite{Flynn:1989iu,Buchalla:1989we} that in the presence of a very heavy top  $Z^0$-penguins, entering $\epe$ with the opposite sign to QCD penguins, cannot be neglected leading to a very strong suppression of $\epe$. 

Therefore, in order
to obtain a useful prediction for $\epe$, the relevant contributions of the QCD
penguin and electroweak penguin operators must be know accurately.
Reviews on $\epe$ can be found in 
\cite{Bertolini:1998vd,Buras:2003zz,Pich:2004ee,Cirigliano:2011ny,Bertolini:2012pu}. See also recent review in \cite{Gisbert:2017vvj} which discusses 
$\epe$ mainly within a chiral perturbative framework including also some large $N$ ideas but having nothing to do with DQCD and reaching very different conclusions  than those presented below.

As far as short-distance contributions (Wilson coefficients of QCD and 
electroweak penguin operators) are concerned, they have been known already
for more than twenty five years at the NLO level
\cite{Buras:1991jm,Buras:1992tc,Buras:1992zv,Ciuchini:1992tj,Buras:1993dy,Ciuchini:1993vr}.
First steps towards the NNLO predictions for $\epe$  have been made in \cite{Buras:1999st,Gorbahn:2004my,Brod:2010mj}. Recently an important progress towards 
the complete NNLO result has been made in \cite{Cerda-Sevilla:2016yzo}. We refer to this paper and the contribution of Maria C{\'e}rda-Sevilla to these proceedings.

The situation with hadronic matrix elements is another story and even if 
significant progress on their evaluation has been made  over the last 25 years,
the present status is far from being satisfactory. In order to describe the 
problem in explicit terms let me write down the NLO formula for $\epe$  
presented in \cite{Buras:2015yba}
\begin{equation}
\frac{\varepsilon'}{\varepsilon} = 10^{-4} \biggl[
\frac{{\rm Im}\lambda_{\rm t}}{1.4\cdot 10^{-4}}\biggr]\!\left[\,a\,
\big(1-\hat\Omega_{\rm eff}\big) \big(-4.1(8) + 24.7\,\bsi\big) + 1.2(1) -
10.4\,\bei \,\right]\,.
\label{AN2015}
\end{equation}
This formula has been obtained by assuming that the real parts of the $K\to\pi\pi$ isospin amplitudes $A_0$ and $A_2$, which exhibit the $\Delta I=1/2$ rule,
are fully described by SM dynamics. Their experimental values are used to
determine to a very good approximation hadronic matrix elements of all
$(V-A)\otimes (V-A)$ operators  \cite{Buras:1993dy}. The first and the third  term in (\ref{AN2015}) summarize these contributions.
In this manner the main uncertainties in $\epe$ reside in the parameters $\bsi$ and $\bei$ which represent the hadronic matrix elements of the $(V-A)\otimes (V+A)$ QCD penguin and
electroweak penguin operators, $Q_6$ and $Q_8$, respectively.  

The parameters  $a$ and $\hat\Omega_{\rm eff}$ summarize  isospin breaking corrections and include  strong isospin
violation $(m_u\neq m_d)$, the correction to the isospin limit coming from
$\Delta I=5/2$ transitions and  electromagnetic corrections. They can be 
extracted from \cite{Cirigliano:2003nn,Cirigliano:2003gt,Bijnens:2004ai} and are given as follows \cite{Buras:2015yba}
\be\label{OM+}
 a=1.017, \qquad
\hat\Omega_{\rm eff} = (14.8\pm 8.0)\times 10^{-2}\,.
\ee
The latter value differs from the one quoted in \cite{Cirigliano:2011ny} but
is equivalent to it as discussed in detail in \cite{Buras:2015yba} after equation (16) in that paper.

The expression (\ref{AN2015}) tells us that a precise  determination of $\bsi$ and $\bei$ in QCD is crucial. First steps in this direction have been made 
30 years ago in \cite{Buras:1985yx,Bardeen:1986vp,Buras:1987wc}  by calculating them analytically in DQCD in the large $N$ limit 
\cite{Buras:1985yx,Bardeen:1986vp,Buras:1987wc} 
\be\label{LN}
\bsi=\bei=1, \qquad {\rm (large~N~Limit)}\,.
\ee
For many years various authors estimated $\bsi$ and $\bei$ in
a number of other large $N$ approaches \cite{Bijnens:2000im,Hambye:2003cy,Bijnens:2001ps} finding $\bsi$ in the ballpark of $3$ and $\bei > 1$. Similar comment applies to $\bei$ in the dispersive approach \cite{Cirigliano:2001qw,Cirigliano:2002jy}. With such values the SM is fully consistent with the data in 
(\ref{EXP}).

The 2015 results from RBC-UKQCD collaboration and DQCD approach 
contradict this picture. Indeed in 2015
significant progress on the values of $\bsi$ and $\bei$ has been 
made by the RBC-UKQCD collaboration, who presented their results on 
the relevant hadronic matrix elements of the operators $Q_6$ \cite{Bai:2015nea}
and $Q_8$ \cite{Blum:2015ywa}. These results imply the following values 
for $\bsi$ and $\bei$ at $\mu=1.53\gev$ \cite{Buras:2015yba,Buras:2015qea}
\be\label{Lbsi}
\bsi=0.57\pm 0.19\,, \qquad \bei= 0.76\pm 0.05\,, \qquad (\mbox{RBC-UKQCD}).
\ee
While the low value of $\bsi$ in (\ref{Lbsi}) is at first sight very 
surprising, a new analysis in DQCD beyond the large $N$ 
limit in (\ref{LN}) \cite{Buras:2015xba} gives strong support to the values in  (\ref{Lbsi}).
In fact, G{\'e}rard and myself demonstrated explicitly the
suppression of both $\bsi$ and $\bei$ below their large-$N$ limit which 
is caused by meson evolution from scales $\ord(m_\pi)$ where (\ref{LN}) is 
valid to scales $\ord(1\gev)$ at which one can compare with lattice results.
The sign of this evolution is such that both 
$\bsi$ and $\bei$ evaluated at $\mu=\ord(1\gev)$ are decreased below unity and 
the suppression of $\bsi$ is stronger than the one of $\bei$. This 
pattern is consistent with the perturbative evolution 
of these parameters above  $\mu=\ord(1\gev)$ \cite{Buras:1993dy}
and implies a smooth matching 
between meson and quark-gluon evolutions. 
Consequently at scales $\mu=\ord(1\gev)$  the 
inequalities
\be\label{NBOUND}
\bsi< \bei < 1 \, \qquad (\mbox{\rm DQCD})
\ee
can be obtained. More specifically we find
\be\label{DQCDBB}
\bsi(m_c)\le 0.60, \qquad B_8^{(3/2)}(m_c)=0.80\pm 0.10
\ee
in agreement with (\ref{Lbsi}).
The result for $\bsi$ is less precise and we cannot exclude values as low as 
$\bsi=0.50$  and as large as $0.70$ but there is a strong indication that $\bsi < \bei$.
For further details, see \cite{Buras:2015xba}. In fact 
as we demonstrated in the case of $K^0-\bar K^0$ matrix elements and summarized briefly above, DQCD even if not precise provided correct pattern of $B_i$ 
values obtained by lattice QCD with much
higher precision than it was possible so far for $\bsi$ and $\bei$. We are therefore
 confident that future more precise lattice calculations will also confirm 
the pattern in (\ref{NBOUND}).

In this context it should be emphasized that in the past values $\bsi=\bei=1.0$ 
have been combined in phenomenological applications with the Wilson coefficients evaluated at scales $\mu=\ord(1\gev)$. The results above show that this is incorrect and the factorization scale is at very low momenta. But to find it out one has to include non-factorizable contributions as done in \cite{Buras:2015xba}
and determine the scale at which they vanish.

Inserting the lattice results in (\ref{Lbsi})  into (\ref{AN2015}) a detailed 
numerical NLO analysis in \cite{Buras:2015yba} gave\footnote{Some authors refer 
to this result as based on DQCD. Even if DQCD would get similar values, the numbers in (\ref{LBGJJ}) are bases of $\bsi$ and $\bei$ from LQCD.}
\be\label{LBGJJ}
   \epe = (1.9 \pm 4.5) \times 10^{-4} \,,
\ee
roughly $3\sigma$ away from the experimental value in (\ref{EXP}). A subsequent  NLO analysis 
in \cite{Kitahara:2016nld} using also hadronic matrix elements from lattice 
QCD confirmed these findings
\begin{align}
  \label{KNT}
  (\epe)_\text{SM} & = (1.1 \pm 5.1) \times 10^{-4},\qquad {\rm (KNT)}\,.
\end{align}
The difference from (\ref{LBGJJ}) is related to a different input but clearly 
these results are consistent with each other.

While these results, based on the hadronic matrix elements from RBC-UKQCD 
lattice collaboration, suggest some evidence for the presence of NP in hadronic $K$ decays,
the large uncertainties in the hadronic matrix elements in question do not yet
preclude that eventually the SM will agree with data. In this context the 
upper bounds from DQCD in (\ref{NBOUND}) 
are important as they give presently the strongest support to the anomaly 
in question, certainly stronger than present lattice results. Indeed
employing the rather precise lattice value for $\bei$ in (\ref{Lbsi}) and 
setting
$\bsi\le\bei=0.76$,
one  finds varying all other input parameters the upper bound
\be\label{BoundBGJJ}
(\epe)_\text{SM}\le (6.0\pm 2.4) \times 10^{-4} \,,
\ee
still $3\,\sigma$  below the experimental data.  

As the bound in (\ref{NBOUND}) plays a significant role in the conclusion that
NP could be at work in $\epe$, let us remind  sceptical readers about 
other successes of DQCD that we discussed above.
Therefore, I strongly
believe that future more precise lattice calculations of $\bsi$ and $\bei$ will confirm the bound in (\ref{NBOUND}) implying that indeed NP contributes 
significantly to $\epe$ unless the error in the  experimental value in (\ref{EXP}) has been underestimated. In fact taking additional information provided below into account my expectation for the SM value of $\epe$ in the SM is:
\be 
(\epe)_\text{SM}=(5\pm 2)\times 10^{-4},\qquad ({\rm my~expectation~for~SM}).
\ee
Therefore, I strongly disagree with 
the SM estimate in \cite{Gisbert:2017vvj}, where the authors using chiral perturbation framework find  $\epe=(15\pm7)\times 10^{-4}$. From my point of view this
paper demonstrates that $\epe$ cannot be predicted reliably within this framework. Indeed within $2\sigma$ one finds on the one hand $\epe=3\times 10^{-3}$ and 
on the other hand $1\times 10^{-4}$. This framework does not
include the meson evolution and it is not surprizing that the resulting central value of $\epe$ obtained by these authors is so large. 

Additional support for the small value of $\epe$ in the SM comes from the recent reconsideration of the role of FSI in $\epe$ 
\cite{Buras:2016fys} and from first NNLO QCD calculations \cite{Cerda-Sevilla:2016yzo} of QCD penguin contributions. It should also be recalled that NNLO corrections to electroweak penguin contributions calculated already in \cite{Buras:1999st} and not included until now
in the numerical results presented above increase the role of electroweak
penguins by roughly $16\%$ decreasing further $\epe$. In this case an effective
central value of $\bei$ from RBC-UKQCD collaboration is increased to
$(\bsi)_\text{eff}=0.88\pm0.06$. But such effects should be included together with all NNLO corrections.

As far as FSI are concerned
  the chiral perturbation theory practitioners, already long time ago, 
put forward the idea
that both the amplitude ${\rm Re}A_0$, governed by the current-current operator  $Q_2-Q_1$ and the $Q_6$ contribution to the ratio $\epe$ could be 
enhanced significantly through FSI in a correlated manner 
\cite{Pallante:1999qf,Pallante:2000hk,Buchler:2001np,Buchler:2001nm,Pallante:2001he} bringing the SM prediction for $\epe$ in the ballpark of experimental 
data \cite{Gisbert:2017vvj}. However, as shown in \cite{Buras:2016fys}  
 beyond the strict large $N$ limit,  FSI are likely to be relevant for the $\Delta I=1/2$  rule, in agreement with \cite{Pallante:1999qf,Pallante:2000hk,Buras:2000kx,Buchler:2001np,Buchler:2001nm,Pallante:2001he}, but much less relevant  for $\epe$. In particular as demonstrated in  \cite{Buras:2016fys}  the 
correlation between the $\Delta I=1/2$ rule and $\epe$ claimed in these 
papers is broken at the $1/N$ level. 
 Let us hope that new result from RBC-UKQCD collaboration will shed 
light on these different views on $\epe$.

While after the completion of NNLO corrections to Wilson coefficients the 
fate of $\epe$ in the SM will  be in the hands of lattice gauge theorists,
 one should not forget all the efforts made by renormalization group experts 
over almost 30 years that allowed to determine the Wilson coefficients of 
the relevant operators precisely. Without such calculations the matching 
of short distance contributions to long distance contributions represented 
by hadronic matrix elements would not be possible and consequently the 
prediction for $\epe$ would be poorly known even if lattice QCD would 
reach satisfactory precision. For a historical account of these NLO and NNLO 
efforts see \cite{Buras:2011we}.

A number of authors investigated 
what kind of NP could give sufficient upward shift in 
$\epe$ and what would then be  implications for $\kpn$ and $\klpn$. 
The summary of these studies can be found in the reviews in 
\cite{Buras:2016qia,Buras:2016bch,Buras:2016egb} so that I will make only 
general comments on them. The up-to-date list of relevant papers is collected
in Table~\ref{eprimeanomaly}. In these models $\epe$ can be enhanced significantly without violating existing constraints. An exception are leptoquark models
 which we will discuss in the final part of this presentation.

\begin{table}[!htb]
\renewcommand{\arraystretch}{1.1}
\begin{center}
\begin{tabular}{|c|c|c|}
\hline
  NP Scenario & References  & Correlations with
\\
\hline \hline
 LHT & \cite{Blanke:2015wba} &  $\klpn$
\\
$Z$-FCNC & \cite{Buras:2015jaq,Bobeth:2017xry,Endo:2016tnu} & 
$\kpn$ and $\klpn$
\\
 $Z^\prime$  & \cite{Buras:2015jaq}, &  $\kpn$, $\klpn$ and $\Delta M_K$
\\
Simplified Models &   \cite{Buras:2015yca} & $\klpn$
\\
 331 Models  & \cite{Buras:2015kwd,Buras:2016dxz} &  $b\to s\ell^+\ell^-$
\\
Vector-Like Quarks & \cite{Bobeth:2016llm} & $\kpn$, $\klpn$ and $\Delta M_K$
\\
Supersymmetry  & \cite{Tanimoto:2016yfy,Kitahara:2016otd,Endo:2016aws,Crivellin:2017gks,Endo:2017ums} &  $\kpn$ and  $\klpn$ 
\\
2-Higgs Doublet Model  & \cite{Chen:2018ytc,Chen:2018vog} &  $\kpn$ and  $\klpn$ 
\\
Right-handed Currents & \cite{Cirigliano:2016yhc,Alioli:2017ces}  &  EDMs
\\
Left-Right Symmetry & \cite{Haba:2018byj} &  EDMs
\\
Leptoquarks  &  \cite{Bobeth:2017ecx} & all rare Kaon decays
\\
\hline
\end{tabular}
\end{center}
\renewcommand{\arraystretch}{1.0}
\caption{Papers studying implications of $\epe$ anomaly. \label{eprimeanomaly}}
\end{table}

We have seen that one of the reasons for a large uncertainty in the SM prediction for $\epe$ was the strong cancellation between QCDP and EWP contributions.
As stressed in  \cite{Buras:2015jaq} beyond the SM, quite generally either 
EWP or QCDP dominate NP contributions and theoretical uncertainties are 
much smaller because no cancellations take place. We refer to  \cite{Buras:2015jaq} for the discussion of this point.

Finally, in all models listed in Table~\ref{eprimeanomaly} only modifications 
of the Wilson coefficients of SM operators by NP contributions have been considered. However, generally, other operators with different Dirac structures, like 
the ones in (\ref{eq:susy}) could be responsible for the observed $\epe$ 
anomaly. To my knowledge the relevant hadronic matrix elements of these operators have never been calculated in QCD. We hope to present the first results for them  in DQCD soon.

On the other hand the $K\to\pi\pi$ matrix element of the chromomagnetic penguin 
operator has been calculated in DQCD \cite{Buras:2018evv} and found 
to be significantly smaller than previously expected in agreement with the earlier lattice 
QCD calculation by the ETM group of related $K\to\pi$ matrix element of 
this operator \cite{Constantinou:2017sgv}.

\boldmath
\subsection{$K_{L,S}\to\mu^+\mu^-$ and $K_L\to\pi^0\ell^+\ell^-$}
\unboldmath
We will be only very brief about these decays. All are subject to LD uncertainties. $K_{L}\to\mu^+\mu^-$ is CP-conserving, while $K_{S}\to\mu^+\mu^-$ is 
CP-violating and $K_L\to\pi^0\ell^+\ell^-$ are dominated by indirect CP-violation. Yet in the presence of NP both $K_{S}\to\mu^+\mu^-$ and
 $K_L\to\pi^0\ell^+\ell^-$ could still be dominated by direct CP violation. In 
any case all three decays constitute in certain models an important constraint
on model parameters. A recent example are leptoquark models in case one 
would like to remove the $\epe$ anomaly with the help of leptoquarks. We will
discuss this in Section~\ref{LQeprime}.

\section{Recent News}\label{sec:3}

\boldmath
\subsection{SMEFT for $Z$ mediated New Physics}
\unboldmath
\subsubsection{Preliminaries}
It is interesting to ask next what would 
be the implications of the $\epe$  anomaly  for 
rare decays $\kpn$ and $\klpn$. This question can only be answered in 
concrete NP scenarios and we have listed a number of papers above where 
such implications have been studied. In particular in \cite{Buras:2015jaq} 
 a number of correlations between $\epe$ and  $\kpn$ and $\klpn$ has
been presented dependently on NP scenario considered.

 Here we will summarize such implications in a  simple scenario with FCNCs appearing already at tree-level 
and being mediated by $Z$ boson exchange. While 
studies of this type have been presented already some time ago
 \cite{Buras:2012jb, Buras:2015yca, Buras:2015jaq} a rather recent analysis 
in \cite{Bobeth:2017xry} in the framework of SMEFT demonstrates that in 
these papers important contributions to $\Delta F=2$ transitions generated by 
renormalization group effects above the electroweak scale have not been included. A related analysis can be found in \cite{Endo:2016tnu}

Let us then see how such simple models look 
from the point of view of the SMEFT framework and how the analyses in 
 \cite{Buras:2012jb, Buras:2015yca, Buras:2015jaq}  are affected by these new
contributions. We will follow here \cite{Bobeth:2017xry} and for our presentation we will recall the $\Delta F=2$ operators
 in the basis of \cite{Buras:2000if}
\begin{align}
  \label{eq:DF2-VLL}
  O_{{\rm VLL}} & 
  = [\bar s \gamma_\mu P_L d][\bar s \gamma^\mu P_L d] \,, &
  O_{{\rm VRR}} & 
  = [\bar{s} \gamma_\mu P_R d][\bar{s} \gamma^\mu P_R d] \,, 
\\
  \label{eq:DF2-LR}
  O_{{\rm LR},1} & 
  = [\bar{s} \gamma_\mu P_L d][\bar{s} \gamma^\mu P_R d] \,, &
  O_{{\rm LR},2} & 
  = [\bar{s} P_L d ][\bar{s} P_R d] \,,
\end{align}
where the summation over colour indices in every current or quark density has 
been made. We show only operators that are relevant in the case of $Z$ exchanges.  Equivalent discussion can be made with the operator basis $\mathcal{O}_i$ 
of \cite{Ciuchini:1997bw} in (\ref{eq:susy}), which we used previously.

The importance of $Z$-mediated FCNC processes has increased recently in view of
the absence of direct NP signals at the LHC. As  the neutral $Z$ is
particularly suited to be a messenger of possible NP even at scales far beyond
the reach of the LHC, the SMEFT framework is very well suited for the proper 
description of the basis structure of such models. In this manner the gauge 
invariance under the SM group can be kept under control and as we will see
renormalization group effects, not only from QCD as done already in
 \cite{Buras:2012jb, Buras:2015yca, Buras:2015jaq}, but also from electroweak 
gauge interactions and  in particular from top Yukawa couplings can be taken 
properly into account \cite{Bobeth:2017xry}. 

\boldmath
\subsubsection{Some Details}
\unboldmath
Let us  then assume that new particles with a common mass $\Lambda$ have been integrated out at some scale
$\muNP \gg \muEW$, giving rise to the SMEFT framework \cite{Buchmuller:1985jz}. The field content of
the SMEFT-Lagrangian are the SM fields and the interactions are invariant under
the SM gauge group. The corresponding Lagrangian can be written as
\begin{align}
  \label{eq:GSM:EFT}
  {\cal L}_{{\rm SMEFT}} & 
  = {\cal L}_{{\rm dim}-4} + \sum_a \Wc{a} \Op{a}\,. 
\end{align}
Here ${\cal L}_{{\rm dim}-4}$ coincides with the SM Lagrangian and a non-redundant
set of operators of dimension six (dim-6), $\Op{a}$, has been classified in 
\cite{Grzadkowski:2010es}. The anomalous dimensions (ADM) necessary for the RG
evolution
from $\muNP$ to $\muEW$ of the SM couplings and the Wilson coefficients $\Wc{a}$ 
are known at one-loop \cite{Jenkins:2013zja, Jenkins:2013wua, Alonso:2013hga}. 
Given some initial coefficients $\Wc{a}(\muNP)$, they can be evolved down to
$\muEW$, thereby resumming leading logarithmic (LLA) effects due to the quartic
Higgs, gauge and Yukawa couplings into $\Wc{a}(\muEW)$. 

It is customary to parametrize FC-quark couplings of the $Z$ as \cite{Buras:2012jb}
\begin{align}  
  \label{eq:Zcouplings}
  \mathcal{L}_{\psi\bar\psi Z}^{\rm NP} & 
  = Z_{\mu} \sum_{\psi = u,d} \bar \psi_i \, \gamma^{\mu} \left( 
        [\Delta_L^{\psi}(Z)]_{ij} \, P_L 
  \,+\, [\Delta_R^{\psi}(Z)]_{ij} \, P_R \right) \psi_j \,, \qquad P_{L,R}=\frac{1}{2}(1\mp \gamma_5)\,
\end{align}
with $[\Delta_{L,R}^{\psi}(Z)]_{ij}$ being complex-valued couplings. We keep
the flavour indices to be arbitrary as the discussion applies not only to
$(ij=sd)$ but also $(ij=bd)$ and $(ij=bs)$ relevant for $B_{s,d}$ systems.

On the other hand  the operators of SMEFT that induce FC quark couplings
to $Z$ are given as follows. The ones with left-handed (LH) quark currents are
\footnote{In order to simplify notations we suppress flavour indices on the operators.} 
\begin{align}
  \label{eq:LH13}
  \Op[(1)]{Hq} & = (H^\dagger i \overleftrightarrow{\cal D}_{\!\!\!\mu} H) 
                   [\bar{q}_L^i \gamma^\mu q_L^j]\,, &
  \Op[(3)]{Hq} & = (H^\dagger i \overleftrightarrow{\cal D}^a_{\!\!\!\mu} H) 
                   [\bar{q}_L^i \sigma^a \gamma^\mu q_L^j]\,.
\end{align}
The ones with right-handed (RH) quark currents are
\begin{align}
  \label{eq:RH1}
  \Op{Hd} & = (H^\dagger i \overleftrightarrow{\cal D}_{\!\!\!\mu} H) 
              [\bar{d}_R^i \gamma^\mu d_R^j], &
  \Op{Hu} & = (H^\dagger i \overleftrightarrow{\cal D}_{\!\!\!\mu} H)
              [\bar{u}_R^i \gamma^\mu u_R^j]\,.
\end{align}
Here $H$ is the Higgs field, $\sigma^a$ are Pauli matrices and ${\cal D}_{\!\mu}$ covariant derivative that 
includes the $W^\pm$ and $Z^0$.

The complex-valued coefficients of these operators are denoted by
\begin{align}\label{SMEFTcoff}
  \wc[(1)]{Hq}{ij}, && \wc[(3)]{Hq}{ij}, && 
  \wc{Hd}{ij}, && \wc{Hu}{ij}.
\end{align}

The $Z$ couplings in (\ref{eq:Zcouplings}) can now be expressed in terms of the latter couplings as follows  \cite{Bobeth:2017xry}
\begin{equation}
  \label{eq:Z-Deltas:dim-6-WC}
\begin{aligned}
  \phantom{x}[\Delta^u_L(Z)]_{ij} & 
  = -\frac{g_Z}{2} v^2 \left[\Wc[(1)]{Hq} - \Wc[(3)]{Hq}\right]_{ij} , &
  [\Delta^u_R(Z)]_{ij} & 
  = -\frac{g_Z}{2} v^2 \wc{Hu}{ij} ,
\\
  [\Delta^d_L(Z)]_{ij} & 
  = -\frac{g_Z}{2} v^2 \left[\Wc[(1)]{Hq} + \Wc[(3)]{Hq}\right]_{ij} , &
  [\Delta^d_R(Z)]_{ij} &
  = -\frac{g_Z}{2} v^2 \wc{Hd}{ij} ,
\end{aligned}
\end{equation}
with $v=246\gev$ being the Higgs vacuum expectation value. 

As $\Wc{a}=\ord(1/\Lambda^2)$, the couplings in (\ref{eq:Z-Deltas:dim-6-WC})
are $\ord(v^2/\Lambda^2)$. If one considers $\Delta F=1$ transitions, the 
leading contributions are just tree-level $Z$ exchanges with one of the vertex
given by (\ref{eq:Zcouplings}) and  (\ref{eq:Z-Deltas:dim-6-WC}) and the 
second flavour conserving vertex being the SM one. Evidently such diagrams
are $\ord(v^2/\Lambda^2)$ and generate  dimension-six contributions
in (\ref{eq:GSM:EFT}).  This is in fact what has been done in \cite{Buras:2012jb, Buras:2015yca, Buras:2015jaq}. So far so good.

But in the latter papers the $\Delta F=2$ processes have been described 
also by simple tree-level $Z$ exchange, this time having on both ends
of the $Z$ propagator the FC vertices in (\ref{eq:Z-Deltas:dim-6-WC}). Evidently
such a contribution is  $\ord(v^4/\Lambda^4)$ and generates one of the dimension-eight contributions in (\ref{eq:GSM:EFT}). While for $\Lambda$ being 
$\ord(1\tev)$ such contributions cannot be neglected, for sufficiently 
large $\Lambda \ge 5\tev$ they cannot compete with dimension-six contributions 
which are $\ord(v^2/\Lambda^2)$. 

The question then arises what are these  dimension-six contributions to 
$\Delta F=2$ processes that represent  $Z$-mediated NP. This question has 
been answered in \cite{Bobeth:2017xry} allowing to identify new effects which have been 
missed in previous literature. These are: 
 
\begin{enumerate}
\item[\bf 1.] 
In the presence of right-handed FC $Z$ couplings, \emph{i.e.} $\mathcal C_{H_d}
\neq 0$ or $[\Delta^d_R(Z)]_{ij}$, inspection of the renormalisation group (RG) equations
due to Yukawa couplings in \cite{Jenkins:2013wua} yields that at $\muEW$ the
left-right $\Delta F=2$ operators $O_{{\rm LR},1}$ in (\ref{eq:DF2-LR}) are generated
and are enhanced by the large leading logarithm $\ln \muNP/\muEW$. Such operators
are known to provide very important contributions to $\Delta F=2$ observables
because of their enhanced hadronic matrix elements and an additional enhancement
from QCD RG effects below $\muEW$, in particular in the $K$-meson system.
As a result \emph{these} operators -- and not $O_{{\rm VRR}}^{ij}$ in (\ref{eq:DF2-VLL}),
as used in \cite{Buras:2012jb, Buras:2015yca, Buras:2015jaq} -- dominate 
$\Delta F=2$ processes. The results in \cite{Jenkins:2013wua} allow the
calculation of this dominant contribution including only leading logarithms 
but this is sufficient for our purposes and even for  scales $\muNP$ as high as $20\tev$ a good approximation is to keep only leading logarithms.

\item[\bf 2.] Because of the usual scale ambiguity present at leading order (LO)
the next-to-leading order (NLO) matching corrections of $\Op{Hd}$
to $\Delta F = 2$ processes at $\muEW$ within SMEFT have to be calculated. One NLO contribution is
obtained by replacing the flavour-diagonal lepton vertex in the SM $Z$-penguin
diagram by $\wc{Hd}{ij}$, which again
generates the operator $O_{{\rm LR},1}^{ij}$ simply because the flavour-changing
part of the SM penguin diagram is LH. In fact this contribution has been first
pointed out in \cite{Endo:2016tnu} and used for phenomenology. Unfortunately,
such contributions are by themselves gauge dependent, simply because the function
$C(x_t)$ present in the SM vertex is gauge dependent. Hence, while the observation
made in \cite{Endo:2016tnu} was important, the analysis of these new contributions
presented there was incomplete\footnote{Meanwhile the authors of  \cite{Endo:2016tnu} included additional contributions and confirmed the results in \cite{Bobeth:2017xry}.}. In \cite{Bobeth:2017xry} the missing
contributions have been calculated using SMEFT, obtaining a gauge-independent contribution. However, the LO contribution
is not only more important due to the large logarithm $\ln\muNP/\muEW$, but has
also opposite sign to the NLO term, allowing to remove the LO scale dependence.  Moreover being strongly enhanced
with respect to the contributions considered in \cite{Buras:2012jb, Buras:2015yca,Buras:2015jaq}, it  has very large impact on the phenomenology; in particular as discussed in detail in \cite{Bobeth:2017xry}  and summarized briefly below
correlations between $\Delta F=2$ and $\Delta F=1$ observables are drastically
changed.

\item[\bf 3.] The situation for LH FC $Z$ couplings is different from the RH
case both qualitatively and quantitatively: inspecting again the RG equations
in \cite{Jenkins:2013wua} one finds that the two operators $\Op[(1)]{Hq}$ and
$\Op[(3)]{Hq}$ in SMEFT listed above generate only the $\Delta F=2$-operator $O_{{\rm VLL}}$ in (\ref{eq:DF2-VLL}) that is
dominant already in the SM. The operator structure is then the same as in \cite{Buras:2012jb}. The resulting NP effects are then much smaller
than in the RH case, because no LR operators are present. But now 
comes an important difference from \cite{Buras:2012jb}.
The
correlations between $\Delta F=1$ and $\Delta F=2$ processes are weakend
very significantly: while $\Delta F=1$ transition amplitudes are proportional to the 
sum $\Wc[(1)]{Hq}{} + \Wc[(3)]{Hq}{}$, the leading RG contribution to $\Delta F=2$
processes is proportional to  $\Wc[(1)]{Hq}{} - \Wc[(3)]{Hq}{}$, that is proportional to $\Delta^u_L(Z)]_{ij}$. The appearance of the $u$-quark coupling in 
a process involving $d$-quarks only, is the consequence of $SU(2)_L$ gauge 
invariance: left-handed up- and down-quark couplings belong to doublets under 
 $SU(2)_L$ symmetry. Consequently we have more free parameters and correlations
between $\Delta F=1$ and $\Delta F=2$ processes are hence only present in specific
scenarios, \emph{e.g.} when the couplings are given in terms of the fundamental parameters of a given 
model that can be determined in other processes. This is in
stark contrast to the contributions considered in \cite{Buras:2012jb, Buras:2015yca, Buras:2015jaq}, where the same couplings enter both classes of
processes and no involvement of specific models was necessary. Of course correlations remain in each sector separately, since both
are governed by two complex couplings, but as previously only one complex 
coupling was present, one needs more observables to determine them model 
independently. Moreover, in models where $\Delta F=2$
and $\Delta F=1$ observables are correlated, the constraints become weaker
allowing for larger NP effects in rare decays.

\item[\bf 4.] Also for the operators $\Op[(1,3)]{Hq}$ the NLO contributions 
to $\Delta F=2$ corresponding to the replacement of the flavour-diagonal lepton
vertex in the SM $Z$-penguin diagram  by
$\Wc[(1,3)]{Hq}$ are gauge dependent.
Including the remaining contributions to remove this gauge dependence one finds
two gauge-independent functions of $x_t$, analogous to $X(x_t)$, $Y(x_t)$ and $Z(x_t)$ known from the SM. Since the NLO contributions are different for
$\Wc[(1)]{Hq}$ and $\Wc[(3)]{Hq}$, it is not just their difference contributing
to $O_{{\rm VLL}}$ anymore, but also their sum. 

\item[\bf 5.] At NLO also new gauge-independent contributions are generated which
are unrelated to tree-level $Z$ exchanges and only proportional to $\Wc[(3)]{Hq}$, 
analogous to the usual box diagrams with $W^\pm$ and quark exchanges. They turn
out to be important for gauge-independence and depend not only on the coefficients
for the quark transition under consideration, but also on additional couplings to
the possible intermediate quarks in the box diagrams. But when the hierarchies 
in CKM elements are taken into account, $\Wc[(3)]{Hq}$ for the quark transition under consideration is the only free entry in this part.
\end{enumerate}

It should be stressed in this context that the contributions to $\Delta F=2$
transitions from FC quark couplings of the $Z$ could be less relevant in NP
scenarios with other sources of $\Delta F=2$ contributions. Most importantly,
$\Delta F=2$ operators could receive a direct contribution at tree-level at
the scale $\muNP$, but also in models where this does not happen $Z$ contributions 
could be subdominant. Examples are models in which the only new particles are
vector-like quarks (VLQs), where box diagrams with VLQ and Higgs exchanges
generate $\Delta F=2$ operators at one-loop level \cite{Ishiwata:2015cga,
Bobeth:2016llm}, which were found in these papers to be larger than the $Z$
contributions at tree-level. However, in \cite{Ishiwata:2015cga} the new effects listed above
have not been included. As shown in \cite{Bobeth:2016llm}
 for right-handed FC $Z$ couplings
these box contributions are dwarfed by the LR operator contributions mentioned
at the begining of our list in Kaon mixing, whereas in $B$-mixing they are
comparable. 

We will now summarize the phenomenological impact of these new effects 
on  the analysis in  \cite{Buras:2015jaq}. To this end we will follow 
the strategy that has been proposed in that paper as 
this will show us where this strategy could still be successful and where 
it has to be modified. The main point of this strategy was the determination 
of FC $Z$ couplings from $\epe$ and $\varepsilon_K$ and to use their values 
 to predict branching ratios for $\kpn$, $\klpn$ and NP contribution to 
$\Delta M_K$. As we will see this strategy is still successful in the case
of RH scenario even if numerical results are rather different 
from those presented in \cite{Buras:2015jaq} because of the contributing 
left-right operators. At first sight this strategy  must be significantly modified in the case of the LH scenario because of an additional coupling present
in $\Delta F=2$ transitions. However, it turns out, as far as 
$\kpn$ and $\klpn$  are concerned, the strategy  in \cite{Buras:2015jaq} remains successful as 
$K_L\to\mu^+\mu^-$ and $\epe$ and not $\varepsilon_K$ are the dominant 
constraints for these two decays in the LH scenario.

It should be emphasized that our critical comments 
about the simplified approach in \cite{Buras:2012jb, Buras:2015yca, Buras:2015jaq} do not apply to $Z^\prime$ models considered in these papers. We will 
discuss these models subsequently.

In the strategy in 
 \cite{Buras:2015jaq}
the central
role is played by $\epe$ and $\varepsilon_K$ for which in the presence of NP
contributions we have
\be\label{GENERAL}
  \frac{\varepsilon'}{\varepsilon}
  = \left(\frac{\varepsilon'}{\varepsilon}\right)^{\rm SM}
  + \left(\frac{\varepsilon'}{\varepsilon}\right)^{\rm NP}\,,
\qquad 
  \varepsilon_K\equiv e^{i\varphi_\eps}\, 
  \left[\varepsilon_K^{\rm SM}+\varepsilon^{\rm NP}_K\right] \,.
\ee

As the size of NP contributions is not precisely known the strategy of
\cite{Buras:2015jaq} is to parametrize this contributions
as
\be\label{deltaeps}
\left(\frac{\varepsilon'}{\varepsilon}\right)^{\rm NP}= \kepe\cdot 10^{-3}, \qquad   0.5\le \kepe \le 1.5
\ee
and 
\be \label{DES}
(\varepsilon_K)^{\rm NP}= \keps\cdot 10^{-3},\qquad 0.1\le \keps \le 0.4 \,.
\ee
The ranges for $\kepe$ and $\keps$ only indicate possible size of NP contributions as argued in \cite{Buras:2015jaq} but can also be treated as free parameters.

\boldmath
\subsubsection{Lessons on  NP  Patterns in $Z$ Scenarios}\label{LessonsZ}
\unboldmath

The summary of the lessons is rather brief. On the other hand the 
presentation in \cite{Buras:2015jaq} is very detailed with numerous analytic
expressions. We stress the differences in numerics due to new contributions 
identified in \cite{Bobeth:2016llm}.

{\bf Lesson 1:}
In the LHS, a given request for the enhancement of $\epe$ determines 
the coupling ${\rm Im} \Delta_{L}^{s d}(Z)$. Similar in the RHS the coupling ${\rm Im} \Delta_{R}^{s d}(Z)$ is determined.

{\bf Lesson 2:}
In LHS there is a direct unique implication of an enhanced $\epe$ on $\klpn$: {\it suppression} of
  $\mathcal{B}(\klpn)$. This property is known from NP scenarios in which 
 NP to $\klpn$ and $\epe$ enters dominantly through the modification of 
$Z$-penguins. The known flavour diagonal $Z$ couplings to quarks and leptons 
and the sign of the matrix element $\langle Q_8\rangle_2$ determines this anticorrelation
which has been verified in all models with only LH flavour-violating $Z$ couplings.

{\bf Lesson 3:} 
The imposition of the $K_L\to\mu^+\mu^-$ constraint in LHS 
determines the range for
${\rm Re} \Delta_{L}^{s d}(Z)$ which with the already fixed ${\rm Im} \Delta_{L}^{s d}(Z)$ would allow to calculate the shifts in $\varepsilon_K$ and $\Delta M_K$ if not for new contributions identified in \cite{Bobeth:2017xry} which were not included in  \cite{Buras:2015jaq}. There it was concluded that
these shifts are very small for $\varepsilon_K$ and negligible for 
 $\Delta M_K$. But this conclusion is not valid in the presence of these
new contributions. Moreover, in concrete models new contributions beyond 
$Z$ exchange are possible.
For instance in VLQ models box diagrams with VLQs 
can indeed provide contributions to  $\varepsilon_K$ and $\Delta M_K$ that 
are larger than coming from tree-level $Z$-exchange provided the masses
of VLQs are far above $3~\tev$ \cite{Ishiwata:2015cga,Bobeth:2016llm}. 
In any case 
 $K_L\to\mu^+\mu^-$ determines the allowed range for ${\rm Re} \Delta_{L}^{s d}(Z)$.

{\bf Lesson 4:}
With fixed ${\rm Im} \Delta_{L}^{s d}(Z)$ and the allowed range 
for ${\rm Re} \Delta_{L}^{s d}(Z)$, the range for  $\mathcal{B}(\kpn)$ 
can be obtained. But in view of uncertainties in the  $K_L\to\mu^+\mu^-$ constraint both an enhancement and a suppression of  $\mathcal{B}(\kpn)$ are possible 
and no specific pattern of correlation between  $\mathcal{B}(\klpn)$ and 
 $\mathcal{B}(\kpn)$ is found. In the absence of a relevant $\varepsilon_K$ 
constraint this is consistent with the general analysis in \cite{Blanke:2009pq}.
$\mathcal{B}(\kpn)$  can be enhanced by a factor of $2$ at most due to 
bound on NP contribution to $K_L\to \mu^+\mu^-$ that hopefully will be
 improved in the future.

{\bf Lesson 5:}
As far as the correlation of $\epe$ with $\klpn$ is concerned analogous pattern is found in RHS, although the numerics is different: suppression of 
$\mathcal{B}(\klpn)$ with increasing $\kepe$ 
But the new contributions from LR operators to $\varepsilon_K$ 
have dramatic impact on the results for $\kpn$ presented in \cite{Buras:2015jaq}. Now not 
 $K_L\to \mu^+\mu^-$   but the constraint from $\varepsilon_K$  determines the allowed  enhancement of  $\mathcal{B}(\kpn)$. While in  \cite{Buras:2015jaq} an enhancement of $\mathcal{B}(\kpn)$  up to a factor of  $5.7$ was possible, now only an enhancement  up to a factor of $1.5$ is possible. 

{\bf Lesson 6:}
In a general $Z$ scenario in which the underling theory contains  all the operators in  (\ref{eq:LH13}) and (\ref{eq:RH1}) and simultaneously dimension-eight LR operators are present the pattern of NP effects can
change relative to LH and RH scenarios because of many parameters involved independently of whether new 
contributions considered in \cite{Bobeth:2017xry} are taken into account or not.
As demonstrated in \cite{Buras:2015jaq}  
the main virtue of the general scenario is the possibility of enhancing 
simultaneously $\epe$, $\varepsilon_K$, $\mathcal{B}(\kpn)$ and $\mathcal{B}(\klpn)$ which is not possible in LHS and RHS. Thus the presence of both 
LH and RH flavour-violating currents is essential for obtaining
simultaneously the enhancements in question when NP is dominated by tree-level 
$Z$ exchanges. We refer to examples in \cite{Buras:2015jaq}.
Then  the main message from this analysis is that in the presence of both LH
and RH new flavour-violating couplings of $Z$ to quarks, large departures from 
SM predictions for $\kpn$ and $\klpn$ are still possible. Similar conclusions 
have been reached in \cite{Endo:2016tnu}.

\boldmath
\subsection{Lessons on  NP  Patterns in $Z^\prime$ Scenarios}\label{LessonsZP}
\unboldmath

$Z^\prime$ models 
exhibit quite different pattern of NP effects in the $K$ meson system than the 
LH and RH $Z$ scenarios. In $Z$ scenarios only electroweak 
penguin (EWP) $Q_8$ and 
$Q_8^\prime$ operators can contribute in an important manner to $\epe$ because of flavour dependent diagonal $Z$ coupling to quarks. But in $Z^\prime$ models the diagonal quark couplings can be flavour universal so that QCD penguin operators (QCDP) ($Q_6,Q_6^\prime$) can dominate NP contributions to $\epe$. Interestingly,
the pattern of NP in rare $K$ decays 
depends on whether NP in $\epe$ is dominated by QCDP  or EWP operators \cite{Buras:2015jaq}. This is in fact a new finding, mainly because nobody studied NP
contributions of QCDP to $\epe$ before.

Another striking difference from $Z$ scenarios, known already from previous 
studies, is the increased importance of the constraints from $\Delta F=2$ 
observables as a simple $Z^\prime$ exchange generates six-dimensional operator 
alone without any interferences with SM contributions that played such an
important role in $Z$ cases.
This has two virtues in the presence of the $\epe$ constraint:
\begin{itemize}
\item
The real parts of the couplings are determined for not too a large $\keps$ from the $\varepsilon_K$ constraint. 
\item
There is a large hierarchy between real and imaginary parts of the  flavour 
violating couplings implied by $\epe$ anomaly in  QCDP and EWP scenarios.
As shown in  \cite{Buras:2015jaq}
in the case of QCDP  imaginary parts dominate over the real ones, while
in the case of EWP this hierarchy is opposite  unless
the $\varepsilon_K$ anomaly is absent. This is related to the fact that 
strong suppression of QCDP to $\epe$ by the factor $1/22$ coming from 
$\Delta I=1/2$ rule  requires a large 
imaginary coupling in order to enhance significantly this ratio. This suppression is absent in the case of EWP and this coupling can be smaller.
\end{itemize}

Because of this important difference in the manner QCDP and EWP enter $\epe$, there are  striking differences in the implications 
for the correlation between $\kpn$ and $\klpn$ in these
two NP  scenarios if significant NP contributions to $\epe$ are required.

We refer to numerous plots in  \cite{Buras:2015jaq}  which show clearly the differences between QCDP and EWP scenarios. More details, in 
particular analytic derivation of all these results, can be found there. We extract from these  results the following lessons:

{\bf Lesson 7:}
In the case of QCDP scenario the correlation between 
$\mathcal{B}(\klpn)$ and $\mathcal{B}(\kpn)$ takes place along the branch 
parallel to the Grossman-Nir (GN) bound. 

{\bf Lesson 8:}
In the EWP scenario the correlation   between 
$\mathcal{B}(\klpn)$ and $\mathcal{B}(\kpn)$  
proceeds away from  this branch for diagonal quark couplings 
$\ord(1)$ if NP in $\varepsilon_K$ is present and it is very different 
from the one of the QCDP case as seen in  the plots in \cite{Buras:2015jaq}
allowing a clear distinction between QCDP and EWP scenarios.

{\bf Lesson 9:}
For fixed values of the neutrino and  diagonal quark couplings in $\epe$ the 
predicted enhancements of $\mathcal{B}(\klpn)$ and $\mathcal{B}(\kpn)$ 
are much larger when NP in QCDP is required to remove the 
$\epe$ anomaly than it is the case of EWP. This is simply related to the fact, 
as mentioned above, that the $\Delta I=1/2$ 
rule suppresses QCDP contributions to $\epe$ so that QCDP operators 
are less efficient in enhancing $\epe$ than EWP operators. Consequently 
 the imaginary parts of the flavour violating couplings 
are required to be larger, implying then larger effects in rare $K$ decays.  
Only 
for the diagonal quark couplings $\ord(10^{-2})$ the requirement of 
shifting upwards $\epe$ implies large effects in $\kpn$ and $\klpn$ in EWP scenario. See  \cite{Buras:2015jaq}  for a detail discussion of this point.

{\bf Lesson 10:}
 In QCDP scenario  $\Delta M_K$ is {\it suppressed} and this 
effect increases with increasing  $M_{Z^\prime}$ whereas in the EWP scenario 
 $\Delta M_K$ is {\it enhanced} and this effect decreases with increasing
 $M_{Z^\prime}$ as long as real couplings dominate.  Already on the basis of this property one could differentiate between 
these two scenarios when the SM prediction for $\Delta M_K$ improves.

In summary assuming that the $\epe$ anomaly will be confirmed by lattice QCD 
 and the results from NA62 and KOPIO for $\kpn$ and $\klpn$  will be available it will be easy to select between various scenarios presented above.

\boldmath
\subsection{Leptoquark models and $\epe$ anomaly}\label{LQeprime}
\unboldmath
We will next turn our attention to leptoquark models and investigate how 
these models confront the  $\epe$-anomaly. We have mentioned already that 
several NP scenarios are able to provide sufficient upward shift in $\epe$ and obtain agreement with experiment. See Table~\ref{eprimeanomaly}. These include in
particular tree-level $Z^\prime$ exchanges with explicit realisation in 331
models \cite{Buras:2015kwd, Buras:2016dxz} or models with tree-level $Z$
exchanges \cite{Bobeth:2017xry, Endo:2016tnu} with explicit realisation in
models with mixing of heavy vector-like fermions with ordinary fermions
\cite{Bobeth:2016llm} and Littlest Higgs model with T-parity
\cite{Blanke:2015wba}. Also simplified $Z^\prime$ scenarios \cite{Buras:2015yca,Buras:2015jaq} and the MSSM \cite{Tanimoto:2016yfy,Kitahara:2016otd,Endo:2016aws,Crivellin:2017gks,Endo:2017ums}  and 2-Higgs doublet models \cite{Chen:2018ytc,Chen:2018vog} are of help here. But the interest in studying LQ models arose not from $\epe$ anomaly but 
from  their ability in the explanations of $B$-physics anomalies with selected
 papers in \cite{Hiller:2017bzc,Dorsner:2017ufx,Crivellin:2017zlb,Buttazzo:2017ixm,Calibbi:2017qbu,DiLuzio:2017vat}.
General information on LQ models can be found in \cite{Davidson:1993qk,Dorsner:2016wpm}. In Table~\ref{tab:Models} we list various LQ models.

Already from the beginning one can expect that the $\epe$ anomaly will be a challenge for those LQ
analyses of $B$-physics anomalies in which all NP couplings have been chosen to
be real and those to the first generation set to zero. It should also be
realised that the anomalies $R(D)$ and $R(D^*)$ although being very significant can still be 
explained in some LQ models through a tree-level LQ exchange. On the other 
hand the $\epe$ anomaly,
being even larger, if the bound on $\epe$ in \cite{Buras:2015xba, Buras:2016fys}
is assumed, can only be addressed in these models at one-loop level.  This shows
that the hinted $\epe$ anomaly is a big challenge for LQ models.

These expectations have been confirmed by a very detailed analysis in  \cite{Bobeth:2017ecx}.
Assuming a mass gap to the electroweak (EW)
scale, the main mechanism for LQs to contribute to $\epe$ turns out to be 
 EW gauge-mixing of
semi-leptonic into non-leptonic operators. 
In  \cite{Bobeth:2017ecx} also  one-loop decoupling for scalar
LQs has been performed, finding that in all models with both left-handed and right-handed LQ couplings, that is  $S_1$, $R_2$, and $V_2$ and $U_1$,
 box-diagrams generate numerically strongly enhanced EW-penguin operators 
$Q_{8}$  and $Q_8^\prime$  already at the LQ scale. This behaviour is rather special for LQs as  in most models  $Q_{8}$  and $Q_8^\prime$ operators cannot 
be generated at high scale even at NLO and are generated only in the the RG running 
to low energy scale from the operators $Q_{7}$  and $Q_7^\prime$, respectively.
A good example is the SM and all NP models discussed by us until now.

Investigating correlations
of $\epe$ with rare Kaon processes $\klpn$, $\kpn$, $\klpll$, $\ksm$,
$\Delta M_K$ and $\eps_K$ one finds then  that even imposing only a moderate
enhancement of $(\epe)_{\rm NP} = 5 \times 10^{-4}$ to explain the current
anomaly, hinted by the Dual QCD approach and RBC-UKQCD lattice QCD calculations,
leads to conflicts with experimental upper bounds on rare Kaon processes. They
exclude all LQ models with only a single coupling as an explanation of the
$\epe$ anomaly and put serious constraints on parameter spaces of the
 models  $S_1$, $R_2$, and $V_2$ and $U_1$ where the box diagrams can in 
principle provide a rescue to LQ models provided both left-handed and right-handed couplings are non-vanishing. However, then the 
presence of left-right operators contributing not only to $\epe$ but
also to $D^0-\bar D^0$ and $K^0-\bar K^0$ mixings requires some fine tuning 
of parameters in order to satisfy all constraints. In the case of $V_2$ and $U_1$ the analysis of box diagrams can only be done in a UV completion.

\begin{table}[!htb]
\renewcommand{\arraystretch}{1.1}
\begin{center}
\begin{tabular}{|c|c|c|}
\hline
  Scalar Leptoquark & $\text{SU(2)}_L$  & Vector Leptoquark
\\
\hline \hline
 $S_1$  & singlet & $U_1$ 
\\
 $\tilde S_1$  & singlet & $\tilde U_1$ 
\\
 $ R_2$  & doublet & $V_2$ 
\\
$\tilde R_2$  & doublet & $\tilde V_2$ 
\\
$S_3$  & triplet & $U_3$ 
\\
\hline
\end{tabular}
\end{center}
\renewcommand{\arraystretch}{1.0}
\caption{Leptoquark models. \label{tab:Models}}
\end{table}

Future improved results on $\kpn$ from the NA62 collaboration, $\klpn$
from the KOTO experiment and $\ksm$ from LHCb will even stronger exhibit the
difficulty of LQ models in explaining the measured $\epe$, in case the $\epe$
anomaly will be confirmed by improved lattice QCD calculations. Hopefully also
improved measurements of $\klpll$ decays will one day help in this context.

The main messages of  \cite{Bobeth:2017ecx} are  then the following ones. If the
future improved lattice calculation will confirm the $\epe$ anomaly at the
level $(\epe)_{\rm NP}\ge 5\times 10^{-4}$ LQs are likely not responsible for it.
But if the $\epe$ anomaly will disappear one day, large NP effects in rare $K$ decays that are still consistent with present bounds will be allowed. The analysis in \cite{Bobeth:2017ecx} is rather involved and we will not present it here.
But it is an excellent arena to practice the 
technology of SMEFT and anybody 
who wants to test her (his) skills in SMEFT should study 
 \cite{Bobeth:2017ecx} in detail.

\section{Outlook}\label{sec:4}
\subsection{Visions}\label{Vision}
Let us begin the final section with a dream about the discovery of NP in 
$\kpn$  and $\klpn$ decays as
\be\label{NA62}
\mathcal{B}(\kpn)= (18.0\pm 4.0)\cdot 10^{-11},\qquad ({\rm NA62},~2019)\,,
\ee

\be\label{KOTO}
\mathcal{B}(\klpn)= (12.0\pm 4.0)\cdot 10^{-11},\qquad ({\rm KOTO},~2021)\,,
\ee
and the confirmation of the $\epe$ anomaly as
\be
\epe= (5\pm 3)\cdot 10^{-4}, \qquad ({\rm RBC-UKQCD}, ~2018).
\ee

Looking at various plots in the literature it is clear that 
such a combination of anomalies would be
truly tantalizing with a big impact on our field. On the other hand if NA62 will find $\kpn$  branching ratio significantly below
15.0 in these units, the claim for NP will be much weaker and we will have
to wait until KOTO measures the branching ratio for $\klpn$. As I already 
stated at several places in this talk I have no doubts that $\epe$ anomaly 
will stay with us but as of today it is hard to predict at which level.

Assuming then that the lattice values of $\bsi$ and $\bei$ will not be modified 
significantly and the $\epe$ anomaly will stay with us with $\kepe=1.0$ the by
now old measurement of $\epe$ 
 will allow to exclude certain scenarios and favour other ones.
But this will also depend on the allowed size of NP in $\varepsilon_K$, 
$\Delta M_K$ and rare $B_{s,d}$ decays. In particular it is crucial that 
the present anomalies in $B$-decays will be  clarified as this will help 
to identify proper flavour symmetry at short distance scales and their 
breakdown. This is also the case of visible tensions between $\Delta M_{s,d}$ 
and $\varepsilon_K$.

\subsection{Open Questions}
There is no doubt that in the coming years $K$ meson physics will strike back, 
in particular through improved estimates of SM predictions for $\epe$, 
$\varepsilon_K$, $\Delta M_K$ and $K_{L,S}\to\mu^+\mu^-$ and through crucial 
measurements of the branching ratios for $\kpn$ and $\klpn$. Correlations 
with other meson systems, lepton flavour physics, electric dipole moments 
and other rare processes should allow us to identify NP at very short distance 
scales \cite{Buras:2013ooa} and we should hope that this physics will also be directly seen at the LHC.

Let us then end our short review by listing most pressing questions for the coming 
years. On the theoretical side we have:
\begin{itemize}
\item
{\bf What is the value of $\kepe$?} that we defined in (\ref{deltaeps}).
 Here the answer will come not only from lattice QCD but also through improved values of the CKM parameters, completion of NNLO QCD corrections and from an improved understanding of FSI and isospin breaking effects.  
 The recent analysis in the large $N$ approach in \cite{Buras:2016fys}
 indicates  that 
 FSI are likely to be relevant for the $\Delta I=1/2$  rule  in agreement with previous 
studies \cite{Pallante:1999qf,Pallante:2000hk,Buras:2000kx,Buchler:2001np,Buchler:2001nm,Pallante:2001he}, but much less relevant for $\epe$. It is important
that other lattice QCD groups calculate $\bsi$ and $\bei$, because at the 
end their values are most important for  $\epe$. But if this anomaly will
 persist, it will be mandatory to calculate hadronic matrix elements of new operators that are absent in the SM. I am confident that in DQCD we will be able 
to calculate them soon.
\item
{\bf What is the value of $\keps$?} Here the reduction of CKM uncertainties and  the  theoretical ones in $\eta_{cc}$ are  most important. But the  analysis in \cite{Blanke:2016bhf} indicates that if no NP is  
present in $\varepsilon_K$, it is expected to be found in $\Delta M_{s,d}$.
\item
{\bf What is the value of $\Delta M_K$ in the SM?} Here lattice QCD should provide useful answers. As pointed out in \cite{Buras:2015jaq} the sign of possible departure from data could help in 
distinguishing between different origins of the $\epe$ anomaly. Moreover, as 
pointed out  in the context of VLQ models in \cite{Bobeth:2016llm},
the knowledge of the allowed 
size of NP contributions to  $\Delta M_K$ will have an impact on NP in
$\kpn$ is these models.
\item
{\bf What are the precise values of $\RE A_2$ and $\RE A_0$?} Again lattice QCD 
will play the crucial role here although the main dynamics behind this rule
has been identified long time ago in the DQCD approach.
\end{itemize}

On the experimental side we have:
\begin{itemize}
\item
{\bf What is $\mathcal{B}(\kpn)$ from NA62?} We should possibly get some 
information already in 2019.
\item
{\bf What is $\mathcal{B}(\klpn)$ from KOTO?} We should know it around the year 2021.
\item
{\bf Do $Z^\prime$ or other new particles like VLQs with masses in the reach of 
the LHC exist?} We could know it already this year.
\end{itemize}

Definitely there are exciting times ahead of us! 
 But in order to distinguish between various NP scenarios and study flavour symmetries and their breakdown, correlations with $B_{s,d}^0-\bar B_{s,d}^0$ mixing observables and  decays like $B_{s,d}\to \mu^+\mu^-$, $B\to K(K^*)\ell^+\ell^-$, $B\to K(K^*)\nu\bar\nu$ and $B\to D(D^*)\tau\nu_\tau$ will be  crucial.

\section*{Acknowledgements}
I would like to thank the organizers of Epiphany 2018 for inviting me to such
an interesting workshop and a very pleasant atmosphare. In addition I appreciate the comments by Marcin Chszaszcz on the present text. I would like to thank 
all my collaborators for wonderful time we spent together. This year particular 
thanks go to Jean-Marc G{\'e}rard, Christoph Bobeth, Martin Jung, David Straub and my new
 impressive collaborator Jason Aebischer.
The research presented here was done and financed in the context of the ERC Advanced
Grant project ``FLAVOUR''(267104) and was partially supported by the DFG
cluster of excellence ``Origin and Structure of the Universe''.

\renewcommand{\refname}{R\lowercase{eferences}}

\addcontentsline{toc}{section}{References}

\bibliographystyle{JHEP}
\bibliography{allrefs}
\end{document}